\documentclass[10pt,journal,compsoc]{IEEEtran}
\IEEEoverridecommandlockouts

\usepackage{cite}
\usepackage{url}
\usepackage{balance}
\usepackage{adjustbox}
\usepackage{subfigure}
\usepackage{tabularx} 
\usepackage{pifont}
\usepackage[utf8x]{inputenc} 

\usepackage{amsmath}
\usepackage{array}
\usepackage{multirow}
\usepackage{rotating}
\usepackage{xcolor}
\PassOptionsToPackage{table,xcdraw}{xcolor}

\newcolumntype{L}[1]{>{\raggedright\let\newline\\\arraybackslash\hspace{0pt}}m{#1}}
\newcolumntype{C}[1]{>{\centering\let\newline\\\arraybackslash\hspace{0pt}}m{#1}}
\newcolumntype{R}[1]{>{\raggedleft\let\newline\\\arraybackslash\hspace{0pt}}m{#1}}

\usepackage{bbding}
\usepackage{dblfloatfix}
\usepackage{color, colortbl}
\usepackage{graphicx}
\usepackage{amssymb}
\usepackage{pifont}
\usepackage{algorithm}
\usepackage[noend]{algpseudocode}
\usepackage[nolist]{acronym}
\usepackage{fancyvrb}
\usepackage{listings}

\usepackage{amsthm}

\newcommand{\ie}{{\em i.e., }}
\newcommand{\eg}{{\em e.g., }}

\ifCLASSINFOpdf
\else
\fi

\begin{document}
\title{IoT Network Security: Requirements, Threats, and Countermeasures}

\author{Ayyoob~Hamza,
			 Hassan~Habibi~Gharakheili,  
			 and~Vijay~Sivaraman
	\IEEEcompsocitemizethanks{
		\IEEEcompsocthanksitem A. Hamza, H. Habibi Gharakheili, and V. Sivaraman are with the School	of Electrical Engineering and Telecommunications, University of New South Wales, Sydney, NSW 2052, Australia (e-mails: m.ahamedhamza@unsw.edu.au, h.habibi@unsw.edu.au, vijay@unsw.edu.au).
		}
}
	

\maketitle
\begin{abstract}
IoT devices are increasingly utilized in critical infrastructure, enterprises, and households. There are several sophisticated cyber-attacks that have been reported and many networks have proven vulnerable to both active and passive attacks by leaking private information, allowing unauthorized access, and being open to denial of service attacks. 

This paper aims firstly, to assist network operators to understand the need for an IoT network security solution, and then secondly, to survey IoT network attack vectors, cyber threats, and countermeasures with a focus on improving the robustness of existing security solutions. Our first contribution highlights viewpoints on IoT security from the perspective of stakeholders such as manufacturers, service providers, consumers, and authorities. We discuss the differences between IoT and IT systems, the need for IoT security solutions, and we highlight the  key components required for IoT network security system architecture. For our second contribution, we  survey the types of IoT attacks by grouping them based on their impact . We discuss various attack techniques, threats, and shortfalls of existing countermeasures with an intention to enable future research into improving IoT network security.

\end{abstract}



\IEEEpeerreviewmaketitle

\section{Introduction}\label{sec:intro}

Internet of Things (IoT) refers to network-connected devices that either can sense or control the environment with minimal or no human intervention. These devices are deployed in various sectors ranging from critical networks, households, smart cities to enterprises, health care, manufacturing, and education \cite{sachs2014}. It has been forecasted by Gartner that more than 25 billion devices will be connected by 2021 \cite{gartnerPredict}. The reason for such widespread adoption of IoT is due to the fact that these devices help to improve our daily life by using the data they collect through  sensors or by being able to influence our environment through actuators.

There are more and more unsecured devices being released to the market \cite{kumar2019all}. Many connected IoT devices can be found on search engines such as Shodan \cite{shodan18},  and sites like Insecam \cite{Insecam18} list publicly accessible and vulnerable devices, such that their vulnerabilities can be exploited at scale. For example, Dyn, a major DNS provider, was subjected to a DDoS attack originating from a large IoT botnet comprising thousands of compromised IP-cameras  \cite{Dyn16}. IoT devices, exposing TCP/UDP ports to arbitrary local endpoints within a home or enterprise, and to remote entities on the wider Internet, can be used by inside and outside attackers to reflect/amplify attacks and to infiltrate.

The security of IoT has become a timely and important topic that requires plausible work from the research community. Therefore, the main focus of this paper is to understand the requirement of network-level security solution for IoT systems, and then to identify the research gap in the existing research on attack detection and mitigation. Our first aim in this paper is to understand the perspectives of IoT security and the role of IoT stakeholders; then we highlight the system components required for an IoT network security architecture. The second contribution is to survey various attack techniques, cyber threats, and counter measures with a focus on highlighting ways to improve the existing attack detection methods.

\section{Perspectives on IoT Security}\label{sec:stakeholders}
In this section we discuss various stakeholders of IoT systems, their roles, and the challenges they face related to IoT security. We then discuss the need for IoT security in traditional network settings, and highlight the requirements for an IoT network security solution.

\subsection[IoT Systems Stakeholders]{IoT Systems Stakeholders: Cybersecurity Challenges and Roles}

Security evaluation must be conducted by all IoT stakeholders (manufacturers, users, authorities, and service providers) to reduce IoT cyber threats. In this section we summarize the perspectives and challenges faced by them.


\textbf{Manufacturers:} It is expected that manufacturers provide firmware integrity, traffic encryption, and follow strict software development principles; yet there are many unsecured devices that have been released to the market  \cite{IoTSnp17}. This is mainly because most of the manufacturers are predominantly incentivized to bring new devices to market as quickly and cost-effectively as possible. Therefore, security is an afterthought, if even thought of at all. 
There are two major security challenges for manufacturers: no incentives for securing their products, and a lack of cybersecurity skills.

\textit{No incentives:} ``Security costs money" -- this provides the reason for manufacturers not taking it seriously, although recently privacy and security of IoTs have slightly become a concern for consumers. According to authors of \cite{emami2019exploring, morgner2019security, gopavaram2019iotmarketplace}, consumers are now willing to pay more for a secured product than for an unsecured device, and this  is influencing their buying decision. Moreover, security is always a cost, but not resolving threats will tarnish brand value. Yet, there are manufacturers not responding to reported threats \cite{bitdefenderAttack} for products that are available
in the market, which demonstrates the level of priority that security enjoys from the perspective of manufacturers.

\textit{Lack of cybersecurity skills:} Manufacturers are expected to have security testing skills to cover the full spectrum of IoT ecosystems, which covers layers such as hardware, network, service integration, and cloud servers. Insufficient security testing and failure to update skills are major concerns in IoT product development. There are many security guidelines \cite{iotsecurityFoundation, owaspFoundation, gsmaAssessment} that have been developed to specify basic security requirements and to ease the process of testing. They include demands that the manufacturer should specify the permitted applications for running on the device and that they should also ensure proper patch management. Yet such guidelines cover only the basic security requirements, while attackers are using more complex and novel techniques (\eg \cite{sivaraman2016smart}) to compromise the devices -- identifying devices' vulnerabilities requires a set of sophisticated skills.

\textbf{Users:} Household consumers, estate managers, and network managers are the users who directly use and/or manage these devices. Ideally, these users follow a strict procurement process during purchasing, and install only secured devices with strong encryption and proper maintenance. Insufficient knowledge, inadequate operational testing, lack of automatic asset management, and limited network monitoring skills are major challenges that users face, resulting in installation of vulnerable devices into their networks.

\textit{Insufficient knowledge:}  Typical household consumers do not have the knowledge nor the tools to test the security posture of a device before procurement, and furthermore, the user manuals of such consumer devices do not provide information about the security features and/or risks \cite{blythe2019security}. To resolve such issues, the authors in \cite{IoTSnp17}  propose a color-coded rating system to indicate the security posture of individual IoT device types, but this approach does not explicitly detail the risk of having a specific vulnerability in a particular deployment; therefore, the authors in \cite{emami2020ask, shen2019iot} propose a security label that quantifies the risks of deploying such devices in a network. Meanwhile, in an organization, devices undergo thorough security testing by their installation teams, yet there are many attacks that have been reported on these devices due to a lack of rigorous testing before acceptance.
Therefore an extensive testing framework is required for such devices. To overcome this issue,  authors of \cite{IoTSnp17} propose a systematic step to evaluate the security posture of IoT devices. Yet, the scope of this paper is limited to confidentiality and availability pillars of security, and does not test the integrity pillar. Assessing the integrity is challenging and requires a custom analysis per each of individual device types.

\textit{Inadequate operational testing:} Devices undergo a thorough qualification process during procurement, yet a simple firmware update can change the security posture of a device and open up new vulnerabilities; therefore it is essential to have a continuous testing framework which can automatically isolate devices, undertaking rigorous assessment without impacting the network. Authors of \cite{pering2018taming} detail the requirements for such an automated testing framework to test devices during operation. They use SDN techniques to isolate the subject device by taking it offline and applying all security tests. Such functionality requires security penetration testing skills; and furthermore,  there is a research gap as to how to apply such solutions in household networks.

\textit{Lack of automatic asset management:}  In legacy IT networks, operators often use monitoring agents, embedded into connected devices, to discover network assets and monitor their activities. IoT devices are typically resource-constrained, and hence cannot accommodate agents; and with a multitude of IoT device types coming online, it remains a challenging task to automatically discover and identify IoT assets using traditional solutions. Without solutions that enable real-time monitoring, attackers can successfully move around a network unchecked and undetected. In addition, today,  large-scale digital infrastructure is typically managed by two entities: the Estate Management (assets) and IT department (network). Such disjoint management of information makes for more difficulty in asset tracking. To keep track of these IoT assets and their network behavior, we require a knowledge repository that can identify a wide range of IoT device types in a given organization and keep track of their behavioral changes.

\textit{Limited monitoring:} There are attacks that can infiltrate or conduct fraudulent activities through the existing network connectivity. For example, attackers can
use the sim cards used in IoT devices that are not intended for calls and messages \cite{telecomAttack}  and
use them to transfer fraudulent data. IoT devices are also used by criminals for exfiltrating data stealthily \cite{fishAttack}. Therefore, it is necessary to monitor IoT activity patterns to discover such criminal activities. This can easily be managed  in a decent-size organization with a team of  network engineers and cyber analysts who are able to understand and mitigate vulnerabilities when the network intrusion detectors report; however, household owners lack such skills. To resolve such issues in home networks, authors of \cite{sivaraman2015network} \cite{haddadi2018siotome} proposed security as a service for the household, which can be managed by an external service provider such as Internet service providers (ISP). Their solution employs software defined networking (SDN) paradigm to monitor and manage home networks from the cloud.

\textbf{Authorities: } Government policymakers, regulatory bodies, and industry alliances are developing standards and guidelines for securing IoT devices \cite{calIoTBill, nistStandard, ukStandard, usiotbill, iotAA}. The IoT Cybersecurity Improvement Act of 2019, was introduced by the US Senate and this proposes a minimal baseline security standard to consider during the procurement of devices for government authorities \cite{usiotbill}. IoT baseline security guidelines have also been proposed by the US National Institute of Standards and Technology (NIST)
\cite{ross2016systems} and the European Union Agency for Network and Information Security (ENISA) \cite{enisaguide}, which covers  IoT system security aspects such as authentication, authorization, and key management. IETF ratified a standard, Manufacturer Usage Description (MUD), to secure IoT at network level, which depends on the manufacturer defining the network behavior of their device in the form of access control lists \cite{ietfMUD18}. These standards play a crucial role in improving  security requirements by ensuring data protection, service continuity, and device security, that eventually leads to an increase in the bar of accepting devices prior to installation \cite{brass2018standardising}. There are four major challenges related to current standards including: (a) Limited focus on security, (b) vagueness, (c) no legacy support, and (d) lack of mandating.

\textit{Limited focus on security:} IoT security applies to a wide spectrum such as hardware, application, network, service integration, and cloud servers. However, current standards are limited to a subset of security aspects (\eg \cite{ietfMUD18,gharakheili2019network} focus on network security, and  \cite{ukStandard} is limited to IoT development and vulnerability disclosure). These standards create a partially trusted environment that could lead to a potentially vulnerable system unless all the standards are covered by the manufacturer.

\textit{Vagueness:}  Recommended IoT security guidelines \cite{DHS16,ross2016systems}  are largely qualitative and subject to human interpretation,  therefore inadequate for automated and rigorous application. Furthermore, guidelines for application of the standard do not clearly specify any concrete requirements, rather explained in a broad statement which would be difficult to apply in real settings during a compliance check process. For example, the Californian government signed the first IoT security law in the United States \cite{calIoTBill}. It requires devices sold in California to have  reasonable security features which is  broadly defined as (1) appropriate to the nature and function of the device, (2) appropriate to the information it may collect, contain, or transmit and (3) designed to protect the device and any information contained therein from unauthorized access, destruction, use, modification, or disclosure. Such broad statements would be difficult for manufacturers to  apply and it would also be challenging for lawmakers to create a standard applicable to all  devices; yet this can be considered as a first step in regulating IoTs \cite{standardComments}.

\textit{No legacy support:} Regulating devices that are already on the market would be a challenging task \cite{iotstandardChallenges}. This is because there are devices that use deprecated services and do not support a firmware updating process. Regulating such devices would be a difficult task. Regulators should also focus on providing alternative solutions to control such devices.

\textit{Lack of mandating: } Existing standards are still being developed and their approval goes through a complex voting process, which leads to a significant amount of delay before being mandated. Furthermore, IoT covers a wide area of applications, and hence it becomes challenging to mandate a standard to all those areas since a security requirement in a home network could vary from one necessary for critical infrastructure. This is the reason for bills such as \cite{usiotbill} being focused on mandating only on the US government agencies \cite{brass2018standardising}.

\textbf{Service providers:}  IoT service providers can be categorized into two types:  platform and signaling service providers. Platform service providers act as
controlling servers that directly interact with the device to control or capture data, and  can be either  cloud or managed services. Signaling service providers are primarily used for network services. DNS, NTP and STUN servers are a few examples of these. It is expected that both types of service providers are up-to-date with security fixes and provide service reliability. However, the major challenge is the lack of ability to quantify the assurance of security from service providers.

\textit{No security assurance:} During development, manufacturers tend to trust popular third-party services, however every service may have security flaws. It is essential to test and verify the services that are consumed. For example the authors of  \cite{zhou2019discovering}, showcased that during the process of registering a device with a smart home IoT platform, the device has to communicate with both a mobile app and its cloud platform. During this process, the server maintains the state of the three entities (device, mobile app, and the server). The  authors were able to spoof these states and showed that the device can be hijacked, substituted for and  accessed without any authorization. Similarly, signaling service providers such as DNS and NTP are inherently vulnerable to spoofing \cite{ariyapperuma2007security}; therefore, manufacturers need to be aware of such vulnerabilities, and at least limit the attack surface by only communicating with these services securely (\eg DNSSEC \cite{ateniese2001new} instead of DNS). These service providers also collect and store device data, which can lead to  privacy issues; therefore, manufacturers should implement best-practice storage policies and perform frequent testing to ensure these policies are complied with (\eg data  retention policy).

\textbf{Summary:} There are many advocates for IoT security, and current approaches are manifold; but there is not any clear pathway to achieving security effectively. It is essential that security to be considered during the entire life-cycle of a device, \ie  from development to installation. Since the network is an integral part of an IoT ecosystem, our primary focus in this paper is on IoT network security.

\subsection{IoT versus IT Network Security} 

Little consideration for security  during the design of a communication network and the Internet, has inherently caused many security threats for IT devices \cite{oppliger1997internet}. For example, simple spoofing on ARP packets can be the cause of a ``man in the middle'' attack. There are many security recommendations introduced to secure a network (\eg DNSSEC \cite{herzberg2013dnssec}), yet such recommendations have proven to be challenging to adopt. Traditional IT networks have faced many security threats, leading to evolution of a number of solutions to secure them (\eg firewall, honey pot, segmentation); and they are applied in different parts of the networks.

With the recent popularity of IoT devices, they now coexist with one another in the same IT networks, which has led to various cyber-attacks (\eg when criminals hack the network of a casino through an insecure Internet-connected fish tank \cite{fishAttack}). In a traditional network, there are three security approaches in place to protect the network, which include: isolation, device-level protection, and network-level protection. In what follows, we discuss the limitations of existing IT network security approaches for protecting IoT devices.

\textbf{Isolation:} Isolation is widely applied in large organizational networks and it is used for reducing the attack surface of a device. There are two network isolation techniques:  physical air-gaping, and  virtually segmenting networks into enclaves and restricting communication paths \cite{viswanathan2009survey}. Air-gaping is applied to IoT devices to isolate their traffic by separating all connected devices into another parallel physical network. Segmenting networks into logical enclaves (\eg VLANs, subnets) and restricting communication paths, would limit the spread of risk during the attack. These techniques require the device to be separated from the non-IoT devices, but there are certain cases when the isolation/gap is bridged (\eg a manager within the organization may request the live stream of a physical surveillance camera located on IoT network to be accessed from their laptop located on IT network). Such bridges lead to vulnerable and insecure IoT devices (\eg cameras or printers) co-existing with the rest of network devices, and eventually compromising the security of the entire enterprise network.

\textbf{Device-level protection:} General-purpose devices have enough computing power and resources to cater for security solutions such as anti-virus or anti-malware to protect the host from attacks, or have a monitoring agent embedded into the host to collect system logs for vulnerability and/or forensic analysis. Such protections cannot be applied to resource-constrained IoT devices with limited memory and computing power. This makes them vulnerable even to the simplest attacks \cite{IoTSnp17}. Therefore, a viable approach for protecting these devices is to have a network-based device monitoring solution that continuously monitors individual devices on the network for any possible threats.

\textbf{Network-level protection:} Firewalls, and intrusion detection systems (IDS)  are commonly used solutions to protect devices in a network. A firewall or an IDS solution requires to either learn devices benign behavior (``good'' entities) or model signatures of attacks on devices (``bad'' entities) and these appliances often employ a specialized hardware to fully inspect individual packets. Further, due to heterogeneous nature of IoT systems and device types, each with their own specific intended behaviors and security vulnerabilities, it becomes quite challenging for the firewall/IDS appliances to distinguish normal from abnormal traffic that could be symptomatic of an attack. Furthermore, with thousands of IoT devices connecting to the network, it is challenging to monitor each individual device using legacy techniques, and doing so,  would lead to an increase in the cost of maintenance (both packet inspection cost and hardware cost).

\textbf{Summary: } Existing IT network security solutions are limited in providing security for IoT devices. Therefore, it is essential to have an IoT specific security solution to protect these devices. In next section, we look into the core components required for an IoT network security solution.

\subsection{Key Components of IoT Network Security Solution}\label{sec:networkcomponents}  
In an IoT network, lack of “complete visibility” is one problem. Many network operators do not fully know what IoT devices are connected to their network. This becomes  critical when one or more IoT devices are compromised on the network. Without solutions that enable real-time monitoring, attackers can successfully move around a network unchecked and undetected. Therefore, for an IoT network security solution architecture, it is essential to consider asset discovery, device security, and system security as key components to provide a complete network security.


\textbf{Asset discovery: } This component is required to provide  network administrators with  visibility into connected IoT assets. IoT asset discovery is the process of knowing the type of individual device in the network. Prior works \cite{meidan2017, sivanathan2018} have employed machine learning techniques to classify IoT devices for asset management. The method in  \cite{meidan2017} employs over 300 attributes (packet-level and flow-level) -- the most influential ones are minimum, median, and average of packet sizes, Time-To-Live (TTL), the ratio of total bytes transmitted and received, and the total number of packets with RST flag reset. Work in \cite{sivanathan2018} proposes to use features with less computation cost at run-time. Existing machine learning-based proposals need to re-train their model when a new device type is added – this limits the usability in terms of not being able to transfer the models across deployments. There are three challenges during IoT asset discovery. They are: unknown type, behind firmware or device being compromised. The authors in \cite{sivanathan2020managing} propose that confidence of  prediction can be a good indicator of whether or not  the device is compromised or the firmware is different. Yet for network administrators, it is necessary to have an explainable model that provides the reason for such changes. Therefore, we propose an identification approach that uses SDN techniques to build the network profile of a device and uses the MUD profile as a reference; then we use similarity metrics to identify the type of  device. We have shown in \cite{hamza2019verifying}  how this approach can be used for resolving challenges such as unknown type, behind firmware or whether the device is compromised. However, this identification approach comes with two limitations: (a) unbounded delay in identifying devices, and (b) having different types of IoT devices with the same MUD profile (hence inability to successfully identify them). Having indicated the pros and cons of these two approaches, one may want to combine the two as a pathway for future work.

In a real setting, classifying all devices on the network can be computationally expensive especially in a mixed environment; however, as a prior step, recognizing IoT from non-IoT can help to reduce the monitoring cost \cite{sivanathan2020managing, meidan2017profiliot}. Existing IoT identification techniques assume that connected devices are healthy and there is no systematic method identify an IoT device which is already compromised or under attack.

\textbf{Device network security: } This component enables run-time monitoring of individual IoTs. A device profile (network behavior) can be verified prior to deployment through static analysis and then it can be monitored closely to achieve run time security. Pre-deployment security verification is applicable for large scale organizations which contain complex organization policies (\eg SCADA networks). We in  \cite{hamza2019verifying,hamza2018clear} have used MUD as device behavioral profiles and have checked against organizational policy for compliance. For achieving runtime security, there are three techniques:  signature-based threat detection, behavioral profile-based whitelisting, and anomaly detection.

\textit{Signature-based threat detection: } 
Nearly all deployed solutions, including software tools like Bro \cite{Bro1999} and Snort \cite{Roesch1999}, and commercial hardware appliances belong to this category. There are studies that apply signature-based intrusion detection/prevention in SDN environments \cite{Yoon2015, Piggybacking17}. The signature-based approach is not sufficient for addressing the new and growing security issues that come with the proliferation of IoT devices. Attack signatures cannot be developed for a growing number of IoT devices at scale.

\textit{Behavioral profile based whitelisting:} The behavior of a device can be translated to a network policy (\ie access controls). An example proposal for using network policy to secure IoT devices in an SDN environment is found in  \cite{yu2015handling}; however, their policy grammar requires fine-grained access controls that capture the state change (such as smoke sensed or windows opened) associated with IoT devices, which may be infeasible if manufacturers encrypt their sensing data, and undesirable for network operators who do not want to make semantic interpretations of sensing data; furthermore, the proposed theoretical framework in  \cite{yu2015handling} has not been demonstrated in implementation. In HanGuard \cite{hanguard}, the authors propose an access control model to block unauthorized access of IoT devices from mobile devices. This proposed framework is  limited to local network traffic and mobile to IoT device communication. In \cite{jamaral}, the authors propose a specification-based approach for a wireless sensor network and expect the network operator to define the rules. We believe this is too cumbersome for the network operator; MUD inherently proposes a scalable approach, which allows operators to impose a tight set of rules (down to the port level) for each device, thus limiting its communication to only intended traffic flows. MUD specifications can be fed to an IDS to detect observed behavior that is not as
specified, thereby indicating an anomaly or threat \cite{IoTSnP18ids, ranganathan2019soft}. MUD enables  enforcement of a baseline security for IoT devices by isolating exception traffic that does not match the device intended ACEs. However, studies \cite{hamza2019detecting} \cite{afek2019eradicating}  have shown that  covert attacks are still possible.

\textit{Anomaly detection: } Anomaly detection holds promise as a way of detecting new and unknown threats, but despite extensive academic research \cite{Garcia2009}, it has had very limited success in operational environments. The reasons for this are manifold \cite{sommer2010outside}:  ``normal'' network traffic can exhibit much more diversity than expected; obtaining ``ground truth'' of attacks in order to train the classifiers is difficult; evaluating outputs can be difficult due to the lack of appropriate datasets; false positives incur a high cost on network administrators to investigate; and there is often a semantic gap between detection of an anomaly and actionable reports for the network operator.

\textbf{System network security: } A large IoT infrastructure for smart buildings may consist of many subsystems such as HVAC, lighting, access controllers, occupancy sensors, or physical security systems. These subsystems are often managed by a variety of stakeholders from network architect, network engineers, facility management engineers, and cyber security analysts to device manufactures, system integrators, and building managers throughout the life-cycle of the smart building \cite{IoTSFwp2019}. These stakeholders produce different data schema to maintain information about the physical location, network configuration, or security policies of IoT devices. The lack of a common data model is a major challenge in limiting the interoperability and holistic analysis of heterogeneous IoT systems. This has led to many cyber-attacks – for example, the Shodan search engine  \cite{shodan18} has listed publicly exposed building management systems, which allows attackers to penetrate those networks. Current methods for evaluating the security posture of such environments is at best ad-hoc, and enforcement and monitoring of appropriate access controls from outside and within the organization are lacking. However, securing large IoT systems demands a formal model that enables, at design stage, an evaluation of the attack surface exposed by the smart environment, including assessment of firmware updates, breached elements, and organization policy changes on overall security. Also, the model needs to be enforced at run-time, which should include monitoring the communication flows to detect
anomalous patterns. There are two main requirements to achieve this: first a building data model to capture the physical infrastructure and their relationships, and a security verification model that uses these building data for static and run time security verifications.

\textit{Building Data:} Haystack \cite{haystack}, Brick \cite{brick18} and IFC (Industry Foundation Classes)
\cite{bazjanac1999industry} provide constructs to formally define a meta data model to specify sensors, controllers, their location in buildings, and their inter-relationships. For example Brick describes building entities (sensors, equipment, room, floor and so on) and their relationships by abstracting classes and tags, and its hierarchical constructs allow extension of the brick model to express new entities (\eg , Camera can be derived from Sensor). Brick’s expressiveness and ease of adaption further allows it to build a better query processor, and it uses the Resource Description Framework (RDF) syntax to maintain the system ontology. This enables application developers to interact with the ontology using query-based language (\eg SPARQL \cite{sparql}). Such a knowledge representable model can be of benefit for various security applications.

\textit{Verification models:}  Modeling network-wide level security is still in infancy in the research community. This is mainly due to lack of availability of building structure data. Authors of \cite{celik2018soteria, celik2019iotguard,nagendra2019viscr} aimed to detect or resolve conflicts among trigger-and-action- based policies set by network administrators in IoT environments. Work in \cite{nagendra2019viscr} extends trigger-and-action-based policies to support MUD access-control rules and building/floor constructs. We in \cite{mudBrick}  have extended the Brick schema to support network and MUD elements to capture the knowledge representation of IoT system network communications. Using this data model, we showcased two security applications: first,  a static analysis which applies location defined network policies for verifications; and next, a demonstration showing that distributed attack detection can be improved by modeling the device communications systems based on both logical and physical locations.

\section{IoT Network Attacks and Countermeasures}\label{sec:attacs}

\begin{figure}[t!]	
	\centering
	\includegraphics[width=0.85\linewidth]{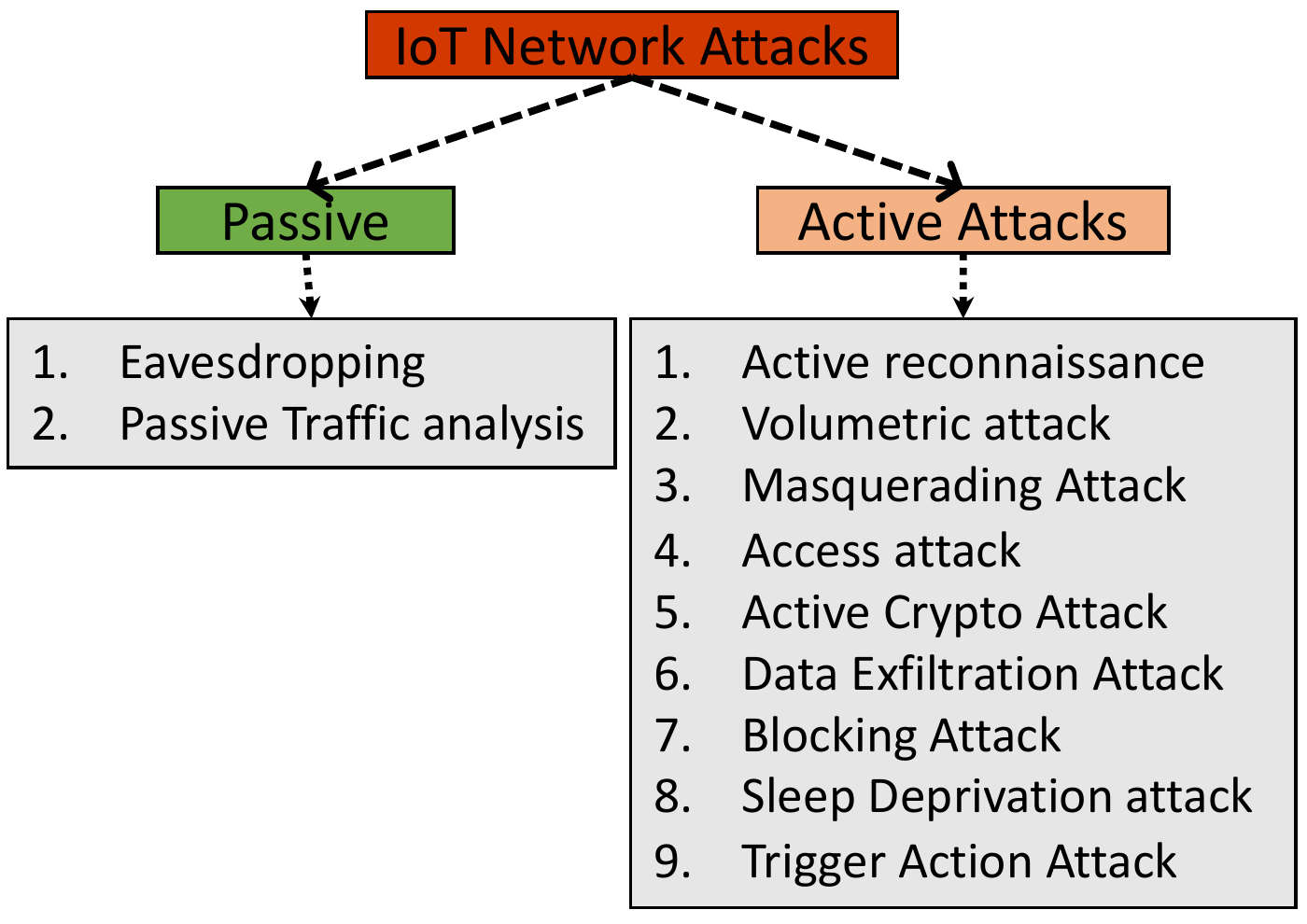}
	\vspace{-2mm}
	\caption{Taxonomy of IoT network attacks.}
	\label{fig:taxonomy}
\end{figure}

In this section, we look into the types of IoT network attacks and then discuss their techniques, implications, and solutions. IoT attacks can be viewed in various dimensions such as impact on information security pillars (\ie confidentiality, integrity, and availability), or based on impact on network layer (\ie application, transport, or data link). However, in this paper we have grouped threats based on the consequences or impact of a potential attack. As Fig.~\ref{fig:taxonomy} indicates, attacks can be categorized into passive or active attacks. IoT network attacks are manifold and we in this paper, have generalized them into 11 types of attacks. There are two types of attacks under passive attacks, and nine under active attacks. Passive attacks don’t leave any network trace by the attacker; rather, they listen to the device communication to collect information about the target. On the other hand, in active attacks, the attacker generates packets directly or indirectly to target the device.

\subsection{Eavesdropping} 
In this attack, the attacker secretly listens to the information that is passed across the devices in the network and this information is used for either profiling the device
or the network.

\textbf{Implications: } Many consumer IoT devices expose themselves by broadcasting device discovery packets such as SSDP and mDNS. These packets contain unique information to determine the type of  device, firmware or device configuration \cite{sivaraman2016smart}. In \cite{sivaraman2016smart}, the authors have shown the implication of such vulnerability by demonstrating that the attacker is able to hijack a device and able to control it remotely using the collected information. Many IoT devices communicate without any encryption in local networks and consider the wireless medium as the last resort of security. However, attackers are launching side-channel attack to access information from these devices. For example, the vulnerability in the Wi-Fi chip has led to broken encryption protocols such as WPA2-Personal and WPA2- Enterprise, which widely affects  many wireless IoT devices and enables the attacker  to take advantage of such vulnerability and launch a side-channel attack to access the device information  \cite{taraseals}. Attackers have used this information to compromise, degrade or disrupt the network.

\textbf{Solutions: } Eavesdropping does not leave any network traces, so it is challenging to detect such attacks. Therefore, the only solution is to prevent such attacks by use of standard cryptography and having firmware upgrade infrastructure in place.

\textit{Use cryptography}:  The only line of practical defense for eavesdropping is cryptography. Manufacturers should ensure that they use the recommended cipher suites for all standard local and Internet communications. Most consumer devices that are allowed to communicate with a mobile application directly in local networks do not use any standard encryption \cite{IoTSnp17} in discovery protocols. Therefore, protocols such as SSDP and mDNS have been misused and have led to many attacks. One solution for device discovery is to use alternative solutions such as having both the device and the application  publish their private IP address to a broker. Implementing such capability brings up an additional cost; therefore, if it is necessary to use discovery protocols, then the device should undergo rigorous testing to ensure that it only publishes  information that does not lead to weak device security.

\textit{Firmware upgrade infrastructure: } This is a standard requirement for any device; side-channel  vulnerabilities are difficult to identify; however, when they are reported, then the manufacturer should have the infrastructure to apply patches through firmware upgrades.

\subsection{Passive Traffic Analysis} 
In this attack, we passively capture the traffic of a device. This can be either within the host network or from outside the network. A rogue router is an example of how such attacks occur. They occur when an attacker can capture the device traffic that passes through and these traffic patterns are then used to deduce the user activity or to profile the device, which can lead to both privacy and physical attacks \cite{acar2018peek}.

\begin{figure}[t!]	
	\centering
	\includegraphics[width=0.75\linewidth]{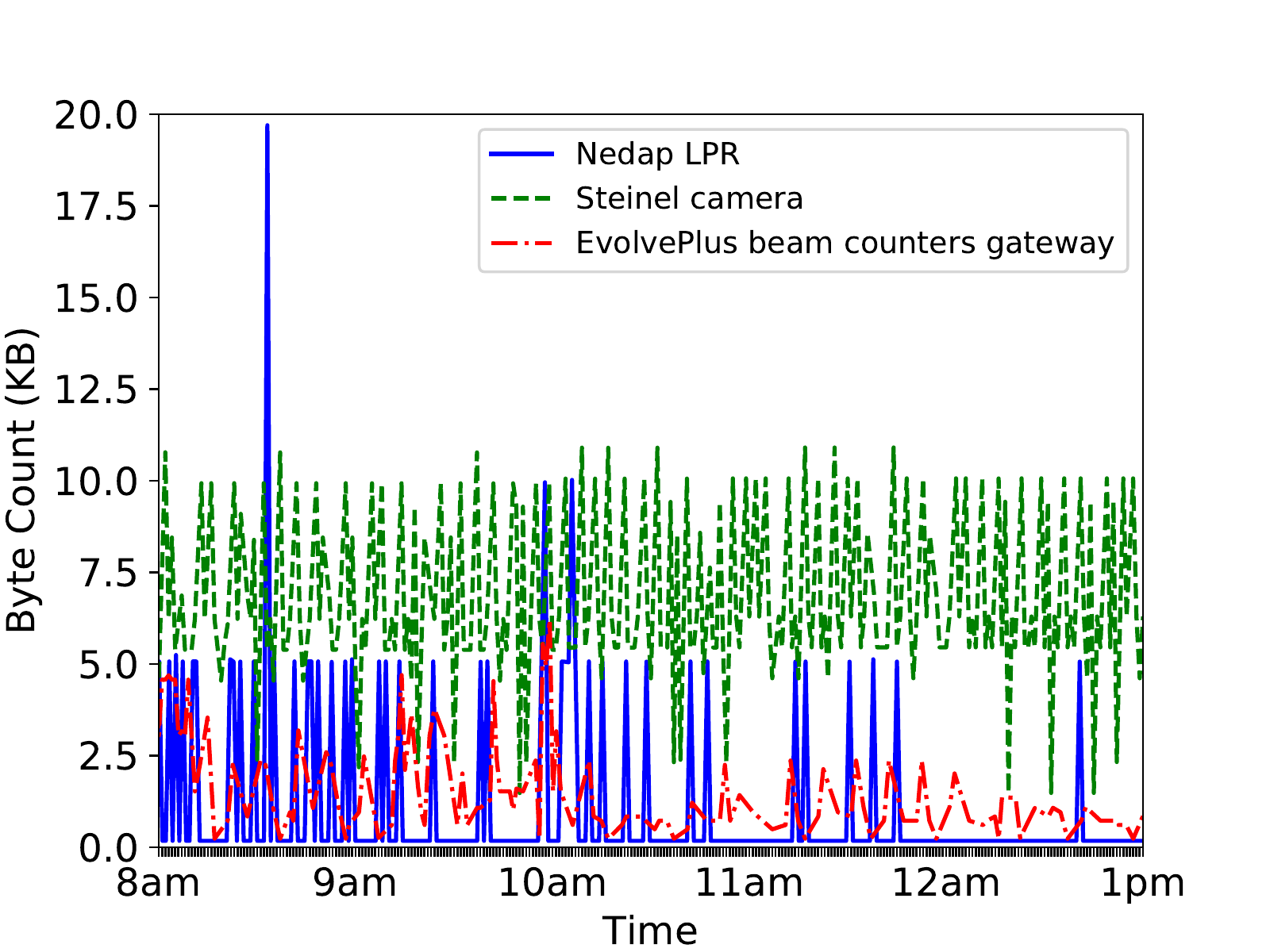}
	\vspace{-2mm}
	\caption{Traffic pattern of representative IoT types.}
	\label{fig:traffic_profile}
\end{figure}

\textbf{Implications:}  IoT device traffic patterns can be categorized into three types  and this is shown in Fig.~\ref{fig:traffic_profile}. They include the device communicating at a fixed interval, communicating only when the user both interacts or communicates with a fixed interval and  interacts. Here the latter two types of traffic patterns capture  user activity. In  \cite{acar2018peek}, authors have demonstrated that even with encrypted traffic they were able to determine the state of a device (\eg turned on or off). In addition, it is also shown that by monitoring IoT traffic on ISP,  in-home activity can still be deduced even if it is encrypted \cite{apthorpe2017spying}. This user activity profiling can be used for a scenario such as theft and it is essential to obfuscate such patterns.

\textbf{Solutions: } There are no network security solutions to detect such an attack; however, there are two preventions to avoid such privacy attacks. They are traffic shaping and tunneling.

\textit{Traffic shaping:} This technique obfuscates all communication so that it all  maps into a similar pattern. The authors in \cite{datta2018developer} developed a python library that  shapes traffic, with low overhead cost; yet this solution would be challenging to apply to a battery-powered device.

\textit{Tunneling:} This technique would aggregate all flows into a single bidirectional flow, which would be challenging to deduce the activity from \cite{apthorpe2017spying} but it does not prevent the attack completely.

\subsection{Active Reconnaissance Attack} 
An attacker probing a device to collect information falls into this category. Reconnaissance attacks can be categorized into two, which are physical and network reconnaissance. Physical reconnaissance requires physical access to the device and there are solutions to limit this by including anti-theft techniques  \cite{tripwire}  such as incorporating an accelerometer to detect  movement of the device. In this paper, we are interested in  network reconnaissance, in which the attacker has to launch an attack through the network to collect information about the device or the network.

\textbf{Implications: } Reconnaissance is harmless to the device functionality, but it can lead to larger attacks on the device; therefore, a network security solution has to identify such attacks before the situation becomes critical. IP address scanning, port scanning or sending an SSDP search query are few attack techniques that have been reported for IoTs \cite{antonakakis2017understanding, shodan18, sivaraman2016smart, acar2018web}. With regard to port scanning or IP address scanning, there are sites like Shodan \cite{shodan18} , which actively probes all the public IP addresses and
their listening port numbers, then it launches attacks to exploit those open ports. Botnets (\eg Mirai botnet) also scan the Internet to identify open telnet ports which in the past, has eventually led to large scale attacks \cite{antonakakis2017understanding}. SSDP search query packets have also been used for fingerprinting devices, which has eventually  led to access attacks \cite{sivaraman2016smart, acar2018web}.

\textbf{Solutions: } To protect from such attacks, the existing network solution uses techniques such as rate limiting, activity modeling, and access controls.

\textit{Rate limiting: } To detect port scanning, the authors in \cite{mehdi2011revisiting} proposed to track new connections and rate limit the tracking; and the authors in \cite{odhaviya2017feasibility} proposed to track TCP RST packets because during scanning, if ports are closed, then the device would respond  with a TCP RST packet, which is an indicator of a scanning attack. Similarly to detecting IP scanning attacks, the Network Security Monitor’s (NSM)  \cite{heberlein1989network}   is able to detect IP scanning if the host connects to more than the configured number of IP addresses. This technique requires thresholding and attackers are now using more sophistical slow port stealth scanning or selective port scanning to avoid getting detected. In addition, such detection techniques also do not work when the attacker selectively sends the SSDP search query to discover the devices.

\textit{Activity modeling: } GrIDS \cite{staniford1996grids} proposes  graph-based activity modeling, which keeps track of all host and  network communication between the hosts. Such a data source provides an individual and holistic view of the network to create more sophisticated rules to detect attacks. Authors in \cite{emerald} proposed a statistical model for a host to capture both short and long term communication One such metric is a volume of syn packets and a sudden increase in volume would be captured as a deviation. Activity modeling can detect scanning attacks but comes with the cost of maintaining the state of the whole network.

\textit{Access controls: } This can be applied to limit the attack surface by using two approaches. The first approach is to allow only devices that are authorized to interact; and next, to allow only the services that the devices are intended to interact with. Authors in \cite{hanguard}  proposed an SDN based architecture that requires the consumers to authorize their devices before they can interact with each other. Furthermore, MUD specification \cite{ietfMUD18} restricts communication to the services that are allowed to. Any communication with disallowed services is considered a threat  \cite{IoTSnP18ids}. However, attacks that are intended  for the device allowed service ports, cannot be detected.

\subsection{Volumetric Attack}
Recent reports \cite{f5Labs17}  show that attackers continue to exploit insecure IoT devices to launch volumetric attacks in the form of DoS, DDoS, brute force, and TCP SYN/UDP flooding. Moreover, the progression of botnets \cite{CiscoReport17, CiscoReport18} such as Mirai and Persirai, infecting millions of IoT devices, is enabling destructive cyber-campaigns of unprecedented magnitude to be launched.

\textbf{Implications: } Mirai botnet was the first,  most notable IoT attack and brought  down many Internet services, which showed the seriousness of vulnerable IoT devices. There are many other botnets that have been discovered over time,  including Persirai \cite{perisirai}, Bashlite \cite{bashlite}, Okiru \cite{okiru}, wicked \cite{wicked}, and Torii \cite{torii}. These botnet attacks require an agent that hijacks the devices by accessing them. In addition, there are DDOS attacks that are launched using reflection on IoT devices \cite{Wisec17}. These attacks are targeted at either remote services or the device itself, with the intention of making them inaccessible  \cite{hamza2019detecting}. 

\textbf{Solutions: }
Volumetric attack detection for computer networks has been studied extensively by the research community, and it looks either for \textit{signatures}  of known attacks, or \textit{anomalies} indicative of deviation from normal behavior.

\textit{Signature-based detection:} 
This is commonly adapted techniques in enterprise networks which requires signatures of all reported threats towards IoT. Any attack that matching the signatures (\eg rate of SYN packets ) are considered as threats. This is a very costly approach in terms of packet inspections and less accurate\cite{IoTSnP18ids} towards zero day attacks. We have shown in \cite{hamza2019detecting} that signature-based tools are only able to detect a limited number of volumetric attacks (to IoTs) that are common for general-purpose computers.

This is a commonly adapted technique in enterprise networks which requires signatures of all reported threats to IoT. Any attack  matching the signatures (\eg rate of SYN packets) are considered  threats. This is a very costly approach in terms of packet inspections and less accurate \cite{IoTSnP18ids} when aimed at zero day attacks. We have shown in \cite{hamza2019detecting} that signature-based tools are only able to detect a limited number of volumetric attacks (to IoTs) that are common for
general-purpose computers.

\textit{Anomaly detection:} 
There are many studies that employ either entropy-based \cite{lakhina2005mining, kumar2007distributed, mehdi2011revisiting,giotis2014combining} or machine learning \cite{braga2010lightweight,cui2016sd,tang2016deep} techniques to detect new volumetric attacks in SDN environments. The entropy-based approach is primarily used for detecting types of volumetric attacks that generate a large number of flows. Authors in \cite{lakhina2005mining, kumar2007distributed,giotis2014combining}  use sample entropy of IP address and ports of both source and destination to determine if there is a large variation captured; and if it exceeds the predetermined threshold, then it is raised as an anomaly. In \cite{kumar2007distributed}, \cite{lakhina2005mining} and \cite{giotis2014combining}  the authors have applied this technique to detect attacks in ISP networks, backbone networks and campus networks respectively. Once an anomaly is detected, then identifying the exact flows (\ie either 3 or 4 tuple flows) on a large scale network is challenging since entropy captures a single value for the captured data and it is challenging to pinpoint the cause of the attack unless maintaining all possible states, which is not feasible for a large network. However, we showcase that this can be achieved in our proposed approach in \cite{mudMitigate} due to maintaining the selected flow states of a device. Works in \cite{braga2010lightweight,cui2016sd,tang2016deep}  use two-class classification (\ie benign and attack). This opposes the expectation from the anomaly-based technique, which  needs to flag deviation from normal behavior \cite{sommer2010outside}. Authors of  \cite{braga2010lightweight, cui2016sd} proposed to use features including flow-level stats (\ie packet/byte count and duration), percentage of bidirectional flows, growth rate of unidirectional flows, and growth rate of number of unique ports, for their classifiers. Work in \cite{tang2016deep} employed deep learning algorithms using a similar set of features to classify normal and abnormal traffic. Authors of \cite{bhunia2017dynamic} applied a technique in \cite{braga2010lightweight} to IoT devices. However, their evaluation is limited to simulated traffic in mininet that does not represent the behavior of real IoT devices.  Not being explainable (knows the device is being attacked but does not know the context of the attack) as signature-based detection, high false positives and training the models using normal and anomalous traffic are fundamental concerns of anomaly based detection, we have overcome this in our work in \cite{hamza2019detecting} by combining MUD with one class machine model trained using only benign data to model traffic pattern of individual MUD access control entries.

\subsection{Masquerading Attack}
In this type of attack, the attack is intended to make the packets appear from somewhere  not intended, or they contain data which is not the original data that was sent. In general-purpose devices, such an attack can be minimized due to a user directly interacting with or host protection in the device can limit such attacks, but IoTs become vulnerable to the simplest attacks.

\textbf{Implications: } As in \cite{huang2019iot}, IoT devices use weak ciphers, weak encryption or no encryption at all. Using a simple DNS spoofing attack, the attacker can make the device trust a fake server and that can lead to a data leak which could impact privacy. Moreover, protocols such as NTP are widely used in IoT devices. Spoofing the time in NTP packets can make the device’s certification fail due to an expired certificate, which can render the device  inactive. Similarly routing information attacks can be launched at  the device to take it offline with very few spoofed packets, and attacks such as ARP spoof can be launched to initiate a man in the middle attack. Yet, these attacks are commonly found in a general-purpose device.  However, IoT devices do not have any agent to provide any feedback and it is challenging for  consumers to deduce or identify the reason for such faults.

\textbf{Solutions: } There is no solution for detecting all possible types of masquerading attacks;  we require a specific attack detection strategy for each attack type. There are two techniques to prevent or detect such attacks, which are activity profiling or using secured communication protocols.

\textit{Activity profiling:} Attacks such as ARP or IP spoof can be detected by inspecting packets and profiling the network  \cite{jinhua2013arp, bharti2013review}. For example, to detect ARP spoofing, a local ARP table can be maintained to identify any conflicts. Similarly, an IP spoof can be detected by maintaining the IP and MAC address mapping and any alteration can be triggered as an attack. To detect the NTP time spoofing attacks, NTP packets also can be inspected for any alteration in NTP offset \cite{garofalo2013gps}; however, to detect a slow rate of change in the NTP offset requires maintaining the state of the device. Detecting DNS spoof is also challenging because it requires  knowing all blacklisted
or whitelisted IP addresses \cite{ramesh2014efficacious}. Activity profiling would lead to higher costs due to deep packet inspection in a high-speed network.

\textit{Use secured protocols: } Using secured network protocol can limit the attack surface. For example, DNSSEC can be used to secure the device from DNS spoofing \cite{ateniese2001new}. It is shown in \cite{IoTSnp17}  that none of the devices they experimented with have used DNSSEC. Further NTP spoofing attacks can be limited by the use of NTP version 4. This limits the NTP offset value but still, slow-changing NTP offset can be harmful; yet, this can be further limited by rate-limiting. In addition, there are  no practical alternatives for securing an ARP spoof attack, but in order to reduce the impact of such an attack, it is recommended to use strong cipher suites for all communication.

\subsection{Access Attack } 
Many consumer IoT devices have been reported to have weak authentication in local services, which has led to many replays and unauthorized access attacks \cite{IoTSnp17, replayattack}. In addition, using default passwords has been the cause of many large scale IoT attacks \cite{password, passwordat}.

\textbf{Implications: } Access attack can lead to many threats. The largest botnet attack, Mirai, was instigated because of having open telnet ports in IoT devices. Devices such as Lifx bulb and TP Link bulb having no authentication in the local services, allowed attackers to take control of them by sending control signals by either replaying or recrafting the messages \cite{insideJob}. Replay attack is launched by capturing control signal traffic and then replaying it; and for the recrafting technique, the attacker  needs to know the payload structure to recraft the packet. The payload structure can be identified through the network traffic or from product API. Furthermore, we identified that the Genbolt IP camera has an unencrypted web portal in the device, through which an attacker can access the device firmware and then can update the firmware by enabling the telnet service with root privileges. In addition, any video URL can be fed to Chromecast without any authentication \cite{insideJob}. Most of such attacks have been launched due to weak authentication in local services. The manufacturer allows unsecured
local services in consumer devices believing home routers would provide access restrictions. However, the authors in \cite{sivaraman2016smart} have demonstrated that devices can be accessed through port forwarding, which is enabled by malware installed through a mobile application. In addition, authors in \cite{acar2018web} have shown that the local services can be accessed through web browsers using DNS rebinding attacks. This shows that it is essential to secure the devices and never to rely on the security of network devices.

\textbf{Solutions: } There are two techniques such as access controls and resistance scheme to limit attacks on devices.

\textit{Access controls: } Traditional network access controls can be used to limit only authorized devices, However, to protect attacks originating from the authorized device, the authors in \cite{hanguard} proposed a mobile agent, which limits the access to authorized applications. Such a solution is difficult to apply in real settings because requesting consumers to install a mobile application is challenging. Therefore, more sophisticated models are required to protect devices from such attacks originating from authorized devices. One possible research pathway is to model the user interaction with the authorized devices, yet it is being said it is very challenging to capture the ground data.

\textit{Resistant scheme: } Use password protection for all services. Applying rate-limiting for password retrying and using session-id for individual message transactions \cite{feng2017replay}  would limit the attack surface.

\subsection{Active Crypto  Attack}
Using strong ciphers is the first step to achieving security, yet it is reported in
\cite{huang2019iot} that many IoT devices do not use strong ciphers, and due to this, many attacks on the integrity security pillar have been reported.

\textbf{Implications: } Ciphers are considered  the fundamental requirement of information security pillars. The authors in \cite{wynn2017sexual}  demonstrated attacks on intimacy devices by hijacking the user session by launching a man in the middle attack. Such
attack breaches users’ privacy and it is considered  cyber rape in some states in countries. This attack uses the SSL split tool to break the existing TLS connection and takes advantage of the vulnerability of the client not verifying the trustworthiness of the provided CA certificate. In addition, we have verified that off the shelf consumer devices such as D-Link camera, Joodan camera, and Genbolt IP camera use lack of cryptographic setting, which allows attackers to launch man in the middle attacks to capture images that the device is transmitting towards the mobile app.

\textbf{Solutions: }  The only prevention is to follow proper security settings by using strong and recommended cipher suites, for example, use TLSv2 algorithm for better security.  Making sure the client validates the CA certificate with its trust store, and checking the expiry date of the certificate are the basic steps to achieve secured connections  \cite{ssllabls}. The network security solution can identify whether  the cipher suites that the device is using for communication are strong and recommended, but it cannot verify whether the client is verifying the certificates in runtime, unless launching a fake certificate test by isolating the device.

\subsection{Data Exfiltration Attack }
Exfiltration of sensitive data is a major concern in organization networks. These attacks are common in  general-purpose devices. The attacker extracts sensitive data and passes it through the covert channel to the command and control (CNC) servers. There are many methods to exfiltrate data and HTTP, FTP, DNS, SSH, Email, are a few examples of those \cite{giani2006data}. 

\textbf{Implications: }  Researchers in \cite{do2016data} have demonstrated an attack where they could collect sensitive documents that are printed using 3D printers. Exfiltrated data can lead to ransomware attacks \cite{yaqoob2017rise}. The collected data are usually transmitted through covert channels using common protocols such as DNS, HTTP, and SMTP \cite{mitreExfiltrationattacks}. Interestingly, the authors in \cite{ronen2016extended} demonstrated a novel type of ex filtration attack by
transmitting the data to outside by changing the intensity of smart light and a receiver from outside then decodes the data.

\textbf{Solutions: }  There are many researchers who have focused on detecting exfiltration attacks targeted at  general purpose devices by incorporating machine learning. For example the authors in \cite{ahmed2019monitoring}  were able to detect an ex filtration attack through DNS covert channel by modeling the pattern of a benign domain name; anything that did not match the benign pattern was marked as a threat. Exfiltration attacks can be launched through various covert channels including encrypted channels such as HTTPS, and SFTP, which fact requires more focus from the research community.

\subsection{Blocking Attack} \label{sec:blockingattacks}
Physical damage, jamming or destructing the functionality of the device falls into the blocking attack category. Many IoT platforms provide the functionality to monitor the health of a device, yet there is a disconnection in information propagation from such platforms to network administrators. It is essential to inform network administrators of such attacks to prevent them in the future.

\textbf{Implications: }  The intention of such an attack is to block the device from communicating with the controllers. Since IoT devices are mostly installed and left unattended, it  is not noticeable unless   a health monitoring service detects it. IoT platforms monitor the health of a device from the device’s health notifications or from the device’s published sensor data. The devices that follow the first approach can be vulnerable since an adversarial attacker can selectively allow only the health service communication but block the others. Such blocking attacks can lead to severe consequences if the attacker targets to disable temperature sensors in a data center or disables motion sensors in critical rooms.

\textbf{Solutions: } There are three techniques to identify such blocking attacks, including monitoring physically, by wireless or by using a network-based solution.

\textit{Physical:} This approach is very costly where a human force is required to monitor
the devices frequently to verify if all device functionalities are working. However, it is challenging to build a workforce to detect such attacks.

\textit{Wireless:} Wireless channels are monitored to identify potential jamming frequencies. Jamming attacks are launched by generating noise to impact the signal to noise ratio of the receivers \cite{jammingAttack}. There are two approaches to avoiding such wireless-based attacks The first is to build the IoT hardware to differentiate such noise from the signal ,but it is challenging to detect whether the noise frequency changes frequently \cite{tang2018jamming}, and implementing such a technique is energy exhausting.  The second approach is to use an external noise detector to identify the increase in noise; however, this is costly for  larger IoT deployments.

\textit{Network based:} Any blocking attack has to eventually impact the network communication. In \cite{mudBrick}, we grouped  devices based on  physical location and then we modeled their communication using one class machine learning technique and an attack was triggered if the device network behavior  deviates from the network system behaviour. We demonstrated the first step in detecting an attack by combining physical and network data; however we found that there is  limited research in identifying such threats using network based models.

\subsection{Sleep Deprivation Attack}
For a battery-powered device, the manufacturer intends to extend longevity by in- creasing the sleep state. However, attackers are targeting such devices by preventing them from sleeping.

\textbf{Implications:} The attacker intends to increase the active time by sending frequent packets to process, which eventually leads to battery exhaustion and leads to the device going offline.

\textbf{Solutions:} The frequency of the packets is a deciding factor for the solution. If the frequency   is of a continuous,  high rate, then it would behave as a volumetric attack; however, attackers now use stealth mode by only sending less frequent packets \cite{pirretti2006sleep}. Modeling energy levels and resource usage of a device is an approach to identify such an attack \cite{nash2005towards}, however, such a solution would not notify the network administrators when such an attack occurred. Therefore, network viable solutions are also required. Interestingly, there is a gap in the research that does not focus on network data to identify such attacks. There are three network approaches that can be applied, which are: access control, rate limiting and traffic modeling.

\textit{Access controls: } This approach can be used to limit the attack space by allowing the device that needs to be communicated with; however, attacks from authorized devices will not be detected.

\textit{Rate limiting: } Authors in \cite{hristozov2019protecting}  proposed  rate-limiting to prevent such attack. However, identifying such a single threshold for devices using network data is challenging.

\textit{Traffic modeling: } Modeling the sleep pattern would be an interesting approach and anomaly can be detected if the observed pattern deviates from the norm. In order to model, we need to identify the state of the device but the challenge is to collect data with ground truth of the sleep states.

\subsection{Trigger Action Attack}
Consumer IoT integrations are enabled using trigger and action rules. For example, a fire alarm can have an on or off state, whereas temperature sensors would have  hot, cold, or normal as states. Consumers can create trigger and action rules using these device states. However, attackers are now exploiting such complex event transitions.

\textbf{Implications: } IoTs are expected to communicate with each other to automate function. For example, if the temperature is high, then they turn on the AC and close the window. An attacker can exploit  households to increase the electricity consumption by notifying the AC to turn on while spoofing the states of both windows and temperature. The main cause of such an attack is not being able to verify device
states. In addition, authors in \cite{surbatovich2017some, sapountzis2018ddift} have shown such exploitation in trigger and action rules can lead to DOS and privacy attacks.

\textbf{Solutions: } There are preliminary studies that focus on solving such problems using network data. Such a solution should be able to identify any spoofing. Machine learning is one promising technique to detect such an attack.

\textit{Machine learning models: } The authors in \cite{sivanathan2020managing} proposed a state identification mechanism by  using  machine learners. This study limits the state classification to idle, active and boot. In addition, \cite{apthorpe2017spying} have demonstrated to identify the state motion using the encrypted traffic data \cite{apthorpe2017spying}. This shows the potential of using network data to determine the states. However, identifying multitude of device states for a large number of device types is challenging. It’s due to lack of annotated datasets, and obfuscated communication. There are only limited research studies focus on detecting states using network data, having such a solution can enable to detect the trust of a device.

\section{Surveys on IoT Network Security}\label{sec:prior}

There has been wide range of IoT security surveys undertaken over recent years, and they have focused on various aspects of security in the IoT eco-system \cite{deogirikar2017security,nawir2016internet,abdul2018comprehensive,aufner2019iot, butun2019security, meneghello2019iot,hassija2019survey,ali2019internet,iotxvulnerabilities2020}. The authors in \cite{hassija2019survey} focused on the security of the IoT layers such as sensing, network, middleware, gateway, and application. Studies such as \cite{nawir2016internet, meneghello2019iot}  focused on IoT security based on the domain it is applied to, such as smart home, transportation, health care or wireless sensor networks. The authors in
\cite{abdul2018comprehensive} focused their security analysis on IoT devices based on wireless protocols such as WiFi, NFC, Bluetooth, and Zigbee, And the authors in \cite{iotxvulnerabilities2020} focused their survey on IoT network attacks. While these studies focused on various aspects of security threats, in this paper we focused on IoT network attack, implications and countermeasures.

\section{Conclusion}\label{sec:con}

Vulnerable IoT devices are being deployed and there are many manufacturers  releasing new devices to the market. These devices have been the cause of many cyber attacks, which have led to privacy breach, unauthorized access, or DoS attacks. This paper focused on understanding the role of various stakeholders in the IoT ecosystem and the challenges they face with respect to IoT security; then we focused on the need for IoT network security and its architecture components. We then focused on IoT network security threats, the implications of such threats and countermeasures, which enabled us to identity the threats that require further attention from the research community. We found that there are several research directions regarding this topic that can have a significant impact on  IoT network security and lead to the evolution of novel security solutions.

\bibliographystyle{IEEEtran}
\bibliography{IoTBack}


\vspace{-2cm}

\end{document}